\documentclass[preprint2]{aastex}
\usepackage{natbib}

\newcommand{\kms}{\ifmmode {\rm km\ s}^{-1} \else km s$^{-1}$\fi}
\newcommand{\Msun}{\ifmmode {\rm M}_{\odot} \else M$_{\odot}$\fi}
\newcommand{\Lsun}{\ifmmode {\rm L}_{\odot} \else L$_{\odot}$\fi}
\newcommand{\qo}{\ifmmode q_{\rm o} \else $q_{\rm o}$\fi}
\newcommand{\Ho}{\ifmmode H_{\rm o} \else $H_{\rm o}$\fi}
\newcommand{\ho}{\ifmmode h_{\rm o} \else $h_{\rm o}$\fi}

\newcommand{\vFWHM}{\ifmmode v_{\mbox{\tiny FWHM}} \else
                    $v_{\mbox{\tiny FWHM}}$\fi}
\newcommand{\CCF}{\ifmmode F_{\it CCF} \else $F_{\it CCF}$\fi}
\newcommand{\ACF}{\ifmmode F_{\it ACF} \else $F_{\it ACF}$\fi}
\newcommand{\Halpha}{\ifmmode {\rm H}\alpha \else H$\alpha$\fi}
\newcommand{\Hbeta}{\ifmmode {\rm H}\beta \else H$\beta$\fi}
\newcommand{\Hgamma}{\ifmmode {\rm H}\gamma \else H$\gamma$\fi}
\newcommand{\Hdelta}{\ifmmode {\rm H}\delta \else H$\delta$\fi}
\newcommand{\Lya}{\ifmmode {\rm Ly}\alpha \else Ly$\alpha$\fi}
\newcommand{\Lyb}{\ifmmode {\rm Ly}\beta \else Ly$\beta$\fi}
\newcommand{\HeI}{\ifmmode {\rm He}\,{\sc i}\,\lambda5876 \else 
	          He\,{\sc i}\,$\lambda5876$\fi}
\newcommand{\HeII}{\ifmmode {\rm He}\,{\sc ii}\,\lambda4686 \else 
	           He\,{\sc ii}\,$\lambda4686$\fi}

\newcommand{\heii}{He\,{\sc ii}}

\newcommand{\feii}{Fe\,{\sc ii}}
\newcommand{\feiii}{Fe\,{\sc iii}}

\newcommand{\ciii}{\ifmmode {\rm C}\,{\sc iii} \else C\,{\sc iii}\fi}
\newcommand{\civ}{\ifmmode {\rm C}\,{\sc iv} \else C\,{\sc iv}\fi}

\newcommand{\niii}{N\,{\sc iii}}
\newcommand{\niv}{N\,{\sc iv}}
\newcommand{\nv}{N\,{\sc v}}

\newcommand{\oiii}{O\,{\sc iii}}

\newcommand{\ovi}{O\,{\sc vi}}

\newcommand{\siiv}{Si\,{\sc iv}}

\slugcomment{To be published in Astrophysical Journal}

\shorttitle{Quasar Elemental Abundances at High Redshifts}
\shortauthors{Dietrich et al.}

\begin{document}

%
%

\title{Quasar Elemental Abundances at High Redshifts}

\author{
M.\,Dietrich
 \altaffilmark{1,2},
F.\,Hamann
 \altaffilmark{1},
J.C.\,Shields
 \altaffilmark{3},
A.\,Constantin
 \altaffilmark{3},
J.\,Heidt
 \altaffilmark{4},
K.\,J\"{a}ger
 \altaffilmark{5},
M.\,Vestergaard
 \altaffilmark{6},
and
S.J.\,Wagner
 \altaffilmark{4},
}
\altaffiltext{1}
{Department of Astronomy, University of Florida, 211 Bryant Space Science 
 Center, Gainesville, FL 32611-2055, USA.}   
\altaffiltext{2}
{current address: Department of Physics and Astronomy, Georgia State 
                  University, One Park Place South SE, Atlanta, GA 30303, USA.}
\altaffiltext{3}
{Department of Physics and Astronomy, Ohio University, Athens,
                  OH 45701, USA.} 
\altaffiltext{4}
{Landessternwarte Heidelberg--K\"{o}nigstuhl, K\"{o}nigstuhl 12, 
 D--69117 Heidelberg, Germany.}
\altaffiltext{5}
{Universit\"{a}tssternwarte G\"{o}ttingen, Geismarlandstra\ss e 11, 
 D--37083 G\"{o}ttingen, Germany.}
\altaffiltext{6}
{Department of Astronomy, The Ohio State University, 140 West 18th Av.,
 Columbus, OH 43210-1173, USA.}
\email{dietrich@chara.gsu.edu}

\begin{abstract}
We examine rest-frame ultraviolet spectra of 70 high redshift quasars 
($z \geq 3.5$) to study the chemical enrichment history of the gas closely 
related to the quasars, and thereby estimate the epoch of first star formation.
The fluxes of several ultraviolet emission lines were investigated within the 
framework of the most recent photoionization models to estimate the
metallicity of the gas associated with the high-z quasars. 
Standard photoionization parameters and the assumption of secondary nitrogen 
enrichment indicate an average abundance of $Z / Z_\odot \simeq 4$ to 5 in 
the line emitting gas.
Assuming a time scale of $\tau _{evol} \simeq 0.5 - 0.8$\,Gyrs for the 
chemical enrichment of the gas, the first major star formation for quasars 
with $z\simeq 4$ should have started at a redshift of $z_f \simeq 6 - 8$, 
corresponding to an age of the universe of several $10^8$\,yrs
($H_o = 65$ km\,s$^{-1}$\,Mpc$^{-1}$, $\Omega _M = 0.3$, 
$\Omega _\Lambda = 0.7$). 
We note that this also appears to be the era of re-ionization of the universe.
Finally, there is some evidence for a positive luminosity -- metallicity 
relation in this high redshift quasar sample.

\end{abstract}
\section{Introduction}

Quasars at high redshift are excellent tools to investigate the formation of 
galaxies and supermassive black holes in the early universe, and to probe the 
physical state of their galactic environment up to early cosmic epochs.
In recent years there is growing evidence that quasar activity and the 
formation of their host galaxies, in particular of massive spheroidal systems, 
are closely related.

The presence of dark massive objects (DMOs) in the center of nearly every 
galaxy with a significant spheroidal component supports models that
connect the formation and evolution of galaxies with quasar activity.
It has been shown that the mass of the DMOs, generally regarded as 
super-massive black holes, is closely correlated with the spheroidal mass of 
the host-galaxy (e.g., Kormendy \& Richstone 1995; Magorrian et al.\,1998;
Richstone et al.\,1998; Gebhardt et al.\,2000; Merritt \& Ferrarese 
2001; Tremaine et al.\,2002).
In the context of galaxy evolution, the conditions that give rise to quasars
and massive black holes will also yield solar or super-solar metallicities on 
time scales shorter than $\sim 1$\,Gyr (Arimoto \& Yoshii 1987; Hamann \& 
Ferland 1993; Gnedin \& Ostriker 1997; Fria\c{c}a \& Terlevich 1998; 
Cen \& Ostriker 1999; Romano et al.\,2002).
Additional strong evidence for the relationship between quasar activity, host 
galaxy formation, and intense star formation episodes is provided by the 
detection of large amounts of dust ($\sim 10^8$\,M$_\odot$) and molecular gas 
($\sim 10^{10}$\,M$_\odot$) measured in high redshift quasars
(Andreani, La\,Franca, \& Cristiani 1993; Isaak et al.\,1994;
 Omont et al.\,1996,\,2001; Carilli et al.\,2000,\,2001).
The co-moving number density of quasars and the cosmic star formation rate
are both at least one order of magnitude larger for epochs with $z\ga 1$ than
in the local universe (e.g., Gallego et al.\,1995; Lilly et al.\,1996; 
Connolly et al.\,1997; Tresse \& Maddox 1998; Steidel et al.\,1999;
Lanzetta et al.\,2002). 
The evolution of the space densities of quasars and galaxies with starburst 
activity are also very similar. Finally, there may be a relation between the 
luminosity functions of normal galaxies and quasars (e.g., Dickinson 1998; 
Pettini et al.\,1998; Boyle \& Terlevich 1998).

Quasars at high redshift are therefore of interest as valuable probes to date 
the beginning of the first star formation episodes in the early universe 
(Hamann \& Ferland 1992,\,1993,\,1999; Dietrich et al.\,1999,\,2000).
Currently, about 350 quasars with redshifts $z \geq 4$ are known 
(e.g., Schneider et al.\,1991a,\,b; Storrie-Lombardi et al.\,1996; Anderson et 
al.\,2001; Djorgovski 2002), and several quasars with $z \ga 5$ and even 
$z > 6$ have been recently discovered 
(Fan et al.\,1999,\,2000a,\,b,\,2001; Stern et al.\,2000; Zheng et al.\,2000; 
 Sharp et al.\,2001; Becker et al.\,2001).
The redshift range $z \ga 4$ corresponds to an epoch when the universe was
less than $\sim 10$\,\%\ of its current age 
(H$_o = 65$ km\,s$^{-1}$ Mpc$^{-1}$, $\Omega _M = 0.3$, 
$\Omega _\Lambda = 0.7$; Carroll, Press, \& Turner 1992). 

The chemical composition of the gas associated with quasars can be 
estimated using the broad emission lines in the ultraviolet spectral range 
(for a review, see Hamann \& Ferland 1999).
Unfortunately, the strengths of prominent metal lines, such as 
\civ $\lambda 1549$, relative to Ly$\alpha$ are not sensitive to the overall
metallicity for $Z \ga 0.1 Z_\odot$ (Hamann \& Ferland 1999).
Shields (1976) proposed that the relative nitrogen abundance could be used
as an indirect metallicity indicator.
Assuming that the secondary nitrogen production, i.e., the synthesis of 
nitrogen from existing carbon and oxygen via CNO burning in intermediate 
mass stars (Tinsley 1980; Henry et al.\,2000), is the dominant source for 
nitrogen, this results in $N/O \propto O/H$ and hence 
$N/H \propto (O/H)^2 \propto Z^2$. 
Observations of HII regions indicate that secondary nitrogen production 
and the $N/O \propto O/H$ scaling dominate when the metallicity is above 
$\sim 1/3$ to $\sim 1/2$ solar (Shields 1976; Pagel \& Edmunds 1981; 
van Zee et al.\,1998; Izotov \& Thuan 1999; Pettini et al.\,2002).
It has been noted (Henry et al. 2000; Kobulnicky \& Skillman 1996) that 
departures from the simple $N/O \propto O/H$ relationship can occur if the 
enrichment is dominated by star formation in discrete bursts. This situation 
leads to time-dependent fluctuations in the N/O ratio because of different 
delays in the stellar release of N and O (and C). However, the overall trend 
for increasing N/O with O/H remains. Moreover, there are no reports, to our
knowledge, of large N/O ratios in metal poor interstellar gas. Large N/O 
abundances are an indicator of high metallicities in any scenario that involves
a well-mixed interstellar medium.

Early investigations of the abundances in broad emission-line region (BELR) 
gas were based on several generally weak inter-combination lines like 
\niv ]$\lambda 1486$, \oiii ]$\lambda 1663$, \niii ]$\lambda 1750$, and
\ciii ]$\lambda 1909$ 
(Shields 1976; Davidson 1977; Baldwin \& Netzer 1978; Osmer 1980; 
 Gaskell et al.\,1981; Uomoto 1984). 
The results of these studies already indicated larger than solar metallicity
for the BELR gas. 
Recent studies of high-redshift quasars ($z\ga 3$) provide evidence for
significantly enhanced metallicities up to several times solar. 
These results obtained by studying the emission line properties of quasars 
(Hamann \& Ferland 1992,\,1993; Ferland et al.\,1996; 
 Dietrich et al.\,1999,\,2002a; Dietrich \& Wilhelm-Erkens 2000; 
 Hamann et al.\,2002; Warner et al.\,2002) have been corroborated by studies
of the intrinsic absorption lines (Petitjean et al.\,1994; 
M\o ller et al.\,1994; Hamann 1997; Pettini 1999).
The derived high chemical abundances require an era of major star formation at
some earlier epoch.\footnote{
The high metallicities cannot be caused by contamination by single stars or
small star clusters. Substantial stellar populations must be involved in the
enrichment because the large masses of gas estimated for the BELR require a 
large mass of heavy elements (Baldwin et al.\,2002).}

Hamann \& Ferland (1992,\,1993) and Ferland et al.\,(1996) show that emission 
line ratios involving \nv $\lambda 1240$ are particularly valuable.
Generally, it is observed that \nv $\lambda 1240$ is stronger than expected in
the spectra of high redshift quasars compared to standard photoionization 
models with solar abundance. 
Assuming nitrogen scales roughly as $Z^2$ at solar and higher metallicities
Hamann \& Ferland (1992,\,1993) suggest \nv $\lambda 1240$/\civ $\lambda 1549$ 
and \nv $\lambda 1240$/\heii$\lambda 1640$ as valuable metallicity indicators. 
Recently, Hamann et al.\,(2002) presented results of a detailed investigation
on the influence of the photoionizing continuum flux and spectral shape,
density, and metallicity on emission line ratios. They further quantified the 
metallicity and 
$N/H \propto Z^2$ dependence of various line ratios, including several weak 
inter-combination lines. They favor 
\niii ]$\lambda 1750$/\oiii ]$\lambda 1663$ and 
\nv $\lambda 1240$/(\ovi $\lambda 1034\, +$ \civ $\lambda 1549$) line ratios
as the most robust indicators to measure the gas chemical composition.

In section 2 we describe the sample of the $z\ga 3.5$ quasars studied here.
In section 3 the results of the analysis of the emission line spectra
are presented. We estimate the elemental abundance of the line emitting gas 
based on several diagnostic emission line ratios (Hamann et al.\,2002).
Using these emission-line ratios, the mean metallicity is
$Z/Z_\odot \simeq 4 \,{\rm to}\, 5$ for the BELR gas of the quasar sample we 
observed. The results are discussed and compared with previous studies 
(e.g., Ferland et al.\,1996; Hamann \& Ferland 1999; 
 Dietrich et al.\,1999,\,2002a; Dietrich \& Wilhelm-Erkens 2000; 
 Hamann et al.\,2002; Warner et al.\,2002) in section 4. 
The chemical composition of the BELR gas provides further evidence that the 
first episodes of major star formation started at a redshift of 
$z_f \simeq 6 \,{\rm to}\, 8$, corresponding to an age of the universe of 
several $10^8$ years. This result is in good agreement with recent model 
predictions relating quasar activity with the formation of massive spheroidal 
systems, e.g., the progenitors of early type galaxies.
In the following we assume $H_o = 65$ km\,s$^{-1}$\,Mpc$^{-1}$, 
$\Omega _M = 0.3$, $\Omega _\Lambda = 0.7$. 

\section{High-Redshift Quasar Sample}
\noindent
We have compiled spectra for a sample of 70 high redshift quasars 
($3.5 \la z \la 5.0$) to study the chemical composition of the BELR gas and 
its implications on the star formation history in quasar host galaxies in the 
early universe.

Most of the quasars at redshift $z\ga 4$ were observed by Constantin et 
al.\,(2002). They recorded high signal-to-noise moderate resolution 
spectra over multiple observing runs at the Multiple Mirror Telescope 
Observatory (MMT) and the W.M.\,Keck Observatory. The spectral wavelength 
range was chosen to cover the redshifted Ly$\alpha $ to \heii $\lambda 1640$ 
emission lines.
Dietrich et al.\,(1999,\,2002a) observed a small sample of 11 quasars with 
$z\ga 4$ using FORS\,1 at the VLT unit telescope {\it Antu} in 1998 and 1999. 
These spectra cover a rest-frame wavelength range of $\sim 850 - 2100$\,\AA . 
This wide range allows measurements of not only Ly$\alpha $ up to
the \heii $\lambda 1640$ emission line, but also \ovi $\lambda 1034$, 
\niii ]$\lambda 1750$, and in some cases even \ciii ]$\lambda 1909$. 
The high redshift quasar sample was complemented by observations which were 
kindly provided by Sargent et al.\,(1988,\,1989), Schneider et al.\,(1991a,b), 
Storrie-Lombardi et al.\,(1996), and Steidel \& Sargent (unpublished).
In particular, the quasar spectra obtained by Storrie-Lombardi et al.\,(1996)
cover the wavelength range containing \ovi $\lambda 1034$ for many of the 
sources observed by Constantin et al.\,(2002). Multiple spectra of the same 
object were combined whenever possible. 

Broad-absorption line quasars (BAL\,QSOs) were excluded from this study, 
although there are indications that their emission line properties do not 
differ from non-BAL quasars (Weymann et al.\,1991).
In Table 1, we list the quasars used in this study together with their 
redshifts, intrinsic continuum luminosity $L_\lambda (1450 {\rm \AA})$
(corrected for galactic extinction; Dietrich et al.\,2002b), the covered 
restframe wavelength range, and the references for the observations.

\section{Analysis}

Comparing mean spectra of the $z\ga 4$ quasars with those at $z\simeq 2$
Constantin et al.\,(2002) found evidence for enhanced \nv $\lambda 1240$ 
emission strength in the high-z quasar spectra indicating high metallicities.
In the following we present quantitative metallicity estimates for each 
individual quasar based on several measured emission line ratios.

The quasar spectra were transformed to their restframe using the redshifts 
listed in Table 1. To determine the redshift for each quasar we fit a Gaussian 
profile to the upper part of the \civ $\lambda 1549$ emission line 
($I_\lambda \geq 50$\,\%\ of the peak intensity). 

To measure integrated emission line fluxes, we corrected each quasar spectrum
for Fe emission and the weak contribution of Balmer continuum emission 
while simultaneously obtaining a power-law fit representing the quasar 
continuum.
The power-law fit, F$_{\nu } \propto \nu ^{\alpha}$, was calculated using 
small spectral regions, each 10 to 20 \AA\ wide, which are free of detectable 
emission lines, at $\lambda \simeq 1290$\,\AA , 1340\,\AA , 1450\,\AA , 
1700\,\AA , 1830\,\AA , and 1960\,\AA , respectively. 
To estimate the contribution of the Balmer continuum and Fe line emission, we 
used our results from the analysis of quasar composite spectra based on a 
large quasar sample of about 750 quasars (Dietrich et al.\,2002b).
For the Balmer continuum emission, we calculated a template spectrum with
T$_e = 15\,000$\,K, $\tau _{BaC}=1.0$, and n$_e = 10^8$\,cm$^{-3}$
following Grandi (1982) and Storey \& Hummer (1995).
The Fe emission was defined by an empirical template spectrum (Vestergaard \& 
Wilkes 2001) and the results of detailed model calculations 
(Verner et al.\,1999). The empirical emission template accounts for both 
\feii\ and \feiii\ emission. The spectral width of the Fe emission features 
was adjusted to the FWHM of the \civ $\lambda 1549$ emission line profile of 
each quasar.
We found that the strength of the Balmer continuum emission and of the Fe 
line emission can be estimated as a fraction of the integrated {\it pseudo
continuum} flux in the wavelength range $1420 - 1470$\,\AA . 
The Balmer continuum emission contributes approximately $1.6\pm 0.7$\,\%\ and
the Fe emission amounts to $1.7\pm 0.4$\,\%\ (Dietrich et al.\,2002b). 
In spite of this small contributions, the estimate of the Fe line emission 
improves especially the measurements of \niii ]$\lambda 1750$ and lines such 
as \heii $\lambda 1640$ near the $\lambda 1600$\,\AA\ feature 
(Laor et al.\,1994; Vestergaard \& Wilkes 2001). 

The \civ $\lambda 1549$ line profile was fitted with a broad and narrow 
Gaussian component. 
The approach to reconstruct the emission line profiles with two Gaussian 
components was chosen to measure the integrated line flux; however, each 
individual component has no physical meaning by itself.
These components have been used as a template to measure 
the other emission line fluxes. 
In using this approach to fit other emission lines, the widths of the 
broad and narrow component were fixed in velocity space while their strengths 
were allowed to vary independently. Furthermore, shifts
in velocity space of each component were restricted to a range
of less than a few 100\,km\,s$^{-1}$ with respect to \civ $\lambda 1549$.
Using the components of the \civ $\lambda 1549$ emission line as templates is 
particularly important for measuring \nv $\lambda 1240$ and 
\heii $\lambda 1640$ because they are blended with other emission lines.
Furthermore, employing the \civ\ line profile as a template provides 
measurements of important weak lines like \niii ]$\lambda 1750$ and 
\niv ]$\lambda 1486$ which are usually difficult to measure.
This template-fitting approach is well justified because \civ $\lambda 1549$, 
\nv $\lambda 1240$, and \heii $\lambda 1640$ are all high ionization lines 
(HIL). 
Figure 1 show typical examples of the deblending of the 
Ly$\alpha 1216$ - \nv $\lambda 1240$, 
\civ $\lambda 1549$ - \heii $\lambda 1640$ - \oiii ]$\lambda 1663$, and
\niii ]$\lambda 1750$ emission line profile complexes, respectively.

The measurement of the \niv ]$\lambda 1486$ emission line flux is severely
affected by the blue wing of the broad \civ\ component. Particularly, for 
quasars with broad emission line profiles, the \niv ]$\lambda 1486$ line tends 
to show a low contrast to the outer part of the \civ $\lambda 1549$ line 
profile (see Dietrich \& Hamann 2003, in prep. for further discussion).
The \ovi $\lambda 1034$ emission line flux has been corrected for Ly$\alpha $ 
forest absorption shortward of the Ly$\alpha$ emission line. For this 
correction we assumed that the quasars show an intrinsic unabsorbed
continuum slope of $\alpha = -1.76$ 
($F_\nu \propto \nu^\alpha$) for $\lambda \leq 1200$\,\AA , following Telfer 
et al.\,(2002). 
Provided that the same fraction of the continuum and \ovi $\lambda 1034$
emission line flux were absorbed, we estimated a corresponding correction 
factor. The integrated \ovi $\lambda 1034$ flux was corrected by this factor, 
which varied from 1.17 to 2.77, with an average of $1.75 \pm 0.42$.

The uncertainties of the flux measurements were estimated from the 
multi-component line fit using the scaled \civ $\lambda 1549$ line profile to 
obtain a minimum $\chi^2$ of the fit. For the stronger lines like 
Ly$\alpha 1216$, \nv $\lambda 1240$, \siiv $\lambda 1402$, 
\civ $\lambda 1549$, and \ciii ]$\lambda 1909$ the errors are of the order of 
$\sim 10$\,\%; for the weaker lines, they are $\sim 20 - 40$\,\%. 
These errors do not take into account the uncertainties introduced by the 
placement of the continuum. While the strength of strong emission lines
like \civ $\lambda 1549$ is not changed significantly, the emission line 
flux of weaker lines like \niv ]$\lambda 1486$ can be underestimated by a 
factor of as much as $\sim 2$ (Dietrich \& Hamann 2003, in prep.).

\section{Results}

We calculated emission line ratios relating nitrogen lines with lines of
helium, oxygen, and carbon to derive the chemical composition of the BELR 
gas at high redshifts.
These line ratios are used to estimate the chemical abundances employing the
results of Hamann et al.\,(2002).
They studied in detail the dependence of emission line ratios from the gas 
metallicity and the shape of the ionizing continuum. For several metallicities
and continuum shapes, they calculated grids of photoionization models spanning
a wide range of density, $n_e$, and continuum strength, $\phi _H$, using 
CLOUDY (Ferland et al.\,1998). 
The metallicity was varied from $Z/Z_\odot = 0.2$ to $10$ with all of the
metals scaled in solar ratios, except $N/O \propto Z$. Three
different input continua were used --- a broken power-law continuum with an
UV bump (Mathews \& Ferland 1987), a single power-law continuum with 
$\alpha = -1$, and a segmented power-law that provides the best match to
recent observations (Zheng et al.\,1997; Laor et al.\,1997). 
For more details of the model calculations, see Hamann et al.\,(2002).

The gas metallicity as provided by each of the emission line ratios is 
presented in Figure 2, assuming a segmented power-law continuum.
The uncertainties of the emission line flux measurements are propagated for
each line ratio and are used to infer an error of each metallicity estimate.
The most reliable line ratios are \niii ]/\oiii ] and \nv/(\ovi $+$\civ).
In particular, \niii ]/\oiii ], \nv /(\ovi $+$\civ ), and 
\nv /\civ\ provide reasonably consistent results.
In spite of some differences, the individual metallicities of all 
line ratios indicate several times solar metallicity for the line emitting 
gas close to the quasars at $z\ga 3.5$.\footnote{Recent investigations of the 
photospheric solar abundance indicate that carbon and oxygen are about 30\,\%\
lower (Holweger 2001; Allende Prieto et al.\,2001,\,2002) than the values 
hitherto generally used as solar abundance (Grevesse \& Sauval 1998), while 
the nitrogen abundance remains nearly unchanged within the errors. Hence, this
would reduce the derived abundances for the high redshift quasars by 
$\sim 30$\,\%.}

Generally, the estimates of the metallicity based on these ratios are nearly 
independent of the shape of the photoionizing continuum (Hamann et al.\,2002). 
For comparison, we also estimated metallicities, using each line ratio,
which were calculated with a hard power-law continuum 
($\alpha = -1.0$) and a SED of the ionizing continuum
as suggested by Mathews \& Ferland (1987).
The derived metallicities differ by less than $\sim 25$\,\%\ in most cases.
The strongest influence of the input continuum shape is for \nv /\heii ,
which amounts to a factor of $\sim 2$ between the two most different continuum
shapes. However, Hamann \& Ferland (1993) and Ferland et al.\,(1996) noted 
that this line ratio provides a firm lower limit on $N/He$ when adopting BELR 
parameters that maximize the predicted \nv /\heii\ line ratio, e.g., a hard 
power-law continuum ($\alpha = -1.0$). 

\subsection{Comparison of \niii ], \niv ], and \nv\ Metallicity Estimates}

Table 2 lists the mean and median metallicities which we derived for each of 
the individual emission line ratios. 
To estimate the uncertainty of the mean metallicities inferred for each line 
ratio, we used the errors of the line ratios which are implied by the 
individual line flux measurements, $\sigma_{obs}$ (see section 3). In
addition, the width of the distribution of the metallicities based on the 
corresponding line ratio for each quasar was taken into account using 
the rms of the mean, i.e., $\sigma _{mean}$ (Bevington 1969). These both
contribution are used in quadrature, i.e., $\sigma = \sqrt{\sigma_{obs}^2 +
\sigma _{mean}^2}$, to estimate the errors of the mean metallicities in
Table 2.

For 35 of the 70 high-redshift quasars we could compare the chemical abundance
based on \niii ]/\oiii ] and the ratios involving \nv $\lambda 1240$ 
(Figure 3). 
Although \nv /\civ, as well as \nv /\ovi\ tend to yield higher metallicities 
than \niii ]/\oiii ], even this latter line ratio indicates several times 
solar metallicity.
The mean metallicities inferred by \nv /\civ, \nv /\ovi\ and \nv /(\ovi$+$\civ)
are only larger by a factor of $\sim 1.7$, $\sim 1.6$, and $\sim 1.6$, 
respectively, than those derived from \niii ]/\oiii ] (Table 2). Only a few 
high redshift quasars differ significantly from the shown trend 
(Figs.\,3a, 3b, and 3c).
Hence, emission line ratios involving \niii ] and \nv\ provide reasonably 
consistent estimates of the gas metallicity for quasars. 
The line ratio \nv /\heii\ shows the largest difference in the metallicity 
estimates in comparison with \niii ]/\oiii ] (Fig.\,3d).  
However, this ratio is particularly sensitive to the uncertain ionizing 
continuum shape (Hamann et al.\,2002).
The metallicities we obtained for \nv /\heii\ using a hard power-law 
continuum agree better with the values provided by \niii ]/\oiii ]. 
The average metallicity based on \niii ]/\oiii] (segmented power-law) and 
\nv /\heii\ based on a hard power-law ($\alpha = -1$) are very similar 
($Z/Z_\odot = 3.6 \pm 0.3$ and $2.9 \pm 0.4$, respectively). 

The \niii ]/\ciii ] ratio tends to indicate lower chemical composition than
the other line ratios (Fig.\,2), i.e., solar to two times solar metallicity. 
However, this line ratio is known to be more dependent on the 
temperature of the gas than \niii ]/\oiii ], and the critical 
densities of \niii ] and \ciii ] differ by a factor of $\sim 2$. Hence,
\niii ]/\ciii ] is not as robust as \niii ]/\oiii ]; emission in \niii ] and
\ciii ] may also occur in spatially different parts of the line emitting 
region with different physical conditions (Hamann et al.\,2002).
In addition, the $C/O$ abundance may exhibit a secondary-like behavior at 
large $O/H$, similar to, but not as sharply rising as, $N/O \propto O/H$
(Henry et al.\,2000).
We estimate that, if this $C/O$ behavior was folded into the calculations of
Hamann et al.\,(2002), the metallicities inferred from \niii ]/\ciii ] would
be $\sim 2$ times larger, while the result based on \nv /\civ\ and 
\nv /(\ovi$+$\civ) would be unchanged (because \civ\ is already 
thermalized and a dominant coolant; see Hamann \& Ferland 1999), suggesting
this effect is probably operating.

Figures 4 and 5 compare the metallicities we calculated based on the 
remaining line ratios. The metallicities derived from the \nv\ ratios agree 
reasonably well with each other.
For several quasars we could also compare the metallicities inferred by 
\niv ]/\oiii ] with the results obtained using flux ratios including 
\niii ] and \nv. In Figure 4d the metallicities given by 
\niv ]/\oiii ] are plotted versus those derived from \nv /\civ. Although the 
scatter is quite large, the average metallicities indicated by each line ratio 
are consistent with $\la 5$\,\%\ with each other.
The comparison of the metallicities calculated using \niv ]/\oiii ] and
\niii ]/\oiii ] are displayed in Figure 5c. The chemical composition indicated
by these line ratios is consistent within a factor of $\la 1.7$.

The \niv ]/\civ\ line ratio indicates in general lower metallicities than the 
other ratios (Figures 2, 5a, 5b, and Table 2).
The tendency toward a significantly lower metallicity based on \niv ]/\civ\ 
was also noted by Shemmer \& Netzer (2002). The average
metallicity indicated by \niv ]/\civ\ is a factor of $\sim 2.5$ lower than
those given by \nv /(\ovi$+$\civ). The physical reason for this difference
is hitherto not understood. 
The metallicity estimates based on \niv ]/\civ\ may appear low in part due to
the difficulty in measuring the line flux of the weak \niv ]$\lambda 1486$ 
emission line. 
In particular, for quasars with broad emission line profiles the 
\niv ]$\lambda 1486$ line is located in the outer wing of the 
\civ $\lambda 1549$ profile. With a typical strength of about $\sim 5$\,\%\ of
the \civ $\lambda 1549$ line flux for solar metallicity, this line can be well
hidden in the outer blue wing of \civ . 
However, a close inspection of the quasar spectra already indicates that not 
enough flux can be assigned to \niv ]$\lambda 1486$ to achieve a \niv ]/\civ\ 
line ratio that is consistent with the high metallicities indicated by the 
other emission line ratios. Another problem with this emission 
line ratio is that it compares an inter-combination 
(semi-forbidden) line to a strong permitted line, whose optical depths differ
by several orders of magnitude. The ratio, therefore, is strongly sensitive
to ambiguities in the radiative transfer, e.g., turbulent line broadening.

In Figures 6a and 6b, the mean metallicity based on the \nv\ line ratios is 
compared with those obtained from line ratios involving inter-combination 
lines. The average metallicity based on \nv\ line ratios indicates higher 
abundances than those inferred from inter-combination line ratios 
($Z/Z_\odot = 6.2 \pm 0.4$ and $3.6 \pm 0.4$, respectively). 

%
Although a nitrogen overabundance is commonly assumed as the cause of \nv\ 
enhancement, scattered Ly$\alpha $ and continuum emission by outflowing broad 
absorption line (BAL) gas clouds would tend to bring the \nv\ and 
intercombination estimates into better agreement. However, it is still an open 
question whether all quasars have BAL type outflow that results in absorption 
lines visible only from certain viewing angles. Several models of an outflowing
BAL wind have been studied (Surdej \& Hutsemeker 1987; Hamann, Korista, \& 
Morris 1993; Krolik \& Voit 1998). Hamann \& Korista (1996) found that only a
fraction of the \nv $\lambda 1240$ emission can be ascribed to scattering in 
the BAL region. They estimate upper limits of $\la 30$\,\%\ for 
\nv $\lambda 1240$ and $\la 10$\,\%\ for \civ $\lambda 1549$. We used these 
upper limits to recalculate the emission line ratios involving \nv\ and \civ . 
The scattered light corrected line ratios indicate chemical compositions which 
are in better agreement with \niii ]$\lambda 1750$/\oiii ]$\lambda 1663$ 
(Table 2). Hence, the main result is still valid that the metallicity of the 
gas closely related to quasars at high redshift is several times solar.

In addition to the mean metallicity based on the individual ratios (Table 2),
we calculated the overall mean metallicity of the high-$z$ quasar sample.
For each quasar an average metallicity was calculated using the estimates
from the line ratios measured in the individual spectra.
Furthermore, we have calculated average metallicities for each quasar based on 
the inter-combinations lines, as well as on the \nv\ line ratios, only.
Table 2 lists the average metallicity calculated using all line ratios 
available for each quasar. These average overall abundances are plotted as 
a function of redshift in Figure 7. The overall metallicity for the 70 high 
redshift quasars is 4 to 5 times solar ($Z/Z_\odot = 5.3 \pm 0.3$). 
It should be kept in mind that all of the metallicity estimates given in this 
paper would be $\sim 30$\,\%\ lower if we used the new solar abundances from 
Holweger (2001) and Allende Prieto, Lambert, \& Asplund (2001,\,2002).

\subsection{The Metallicity -- Luminosity Relationship}

We used our quasar sample to test the potential existence of such a
$L$ -- $Z$ relation at high redshifts.
We calculated the rest-frame continuum luminosity for each quasar in our
sample, L$_\lambda (1450 {\rm \AA })$, at $\lambda = 1450$\,\AA .
The investigated high redshift quasars cover a range of 
$10^{42} \la L_\lambda (1450 {\rm \AA }) \la 10^{45}$ 
erg\,s$^{-1}$\,\AA$^{-1}$.
Most of the quasars show a luminosity of log$\,L_\lambda (1450{\rm \AA }) 
\simeq 43 \,{\rm to}\,44$.
In Figure 8, the average metallicity calculated from all ratios in each
quasar, is shown as a function of intrinsic continuum luminosity 
$L_\lambda (1450{\rm \AA })$ for the quasars of our sample. 
A weak trend of increasing $Z/Z_\odot$ toward higher luminosity is visible.
We calculated a least square fit to the luminosity -- $Z/Z_\odot$ distribution
shown in Figure 8 (dashed line).
The correlation, which shows considerable scatter and spans a small range in
luminosity, is marginally significant, with a 6\,\%\ probability of arising by
chance (70 measurements and correlation coefficient $+0.23$; Bevington 1969). 
Although the correlation is not very 
significant, it provides an additional indication for a luminosity -- 
metallicity (L -- $Z/Z_\odot$) relation for quasars (Hamann \& Ferland 
1993,\,1999). 
We are currently in the process of analyzing a sample of $\sim 750$ quasars in
order to study the evolution of their properties, and find for these objects a
trend of increasing metallicity with increasing luminosity 
(Dietrich \& Hamann 2003).

\section{Discussion}

The high metallicities we derived for the BELR gas of the quasars at $z \ga 4$
provide important information about the preceding star formation epoch that is 
required to enrich the gas. This star formation epoch might also mark the time 
of the formation of the quasar host galaxies.
Several groups have noted that the observed elemental abundances near quasars 
are consistent with normal chemical enrichment in the cores of massive 
galaxies. In particular, one-zone chemical evolution models using with a 
relatively flat IMF (compared to the solar neighborhood) typically achieve 
supersolar gas-phase metallicities in less than $\sim 1$ Gyr (e.g., Hamann 
\& Ferland 1992, 1993; Matteucci 1992; Matteucci \& Padovani 1993; 
Padovani \& Matteucci 1993). Comparable short times scales for attaining metal 
rich gas are predicted by more recent multi-zone models that follow the 
dynamical and chemical evolution of forming galaxies (e.g., Fria\c{c}a \&
Terlevich 1998; Granato et al.\,2001; Romano et al.\,2002). In the
multi-zone models, intense star formation on short time scales occurs only in 
the central region of the quasar host galaxy ($r\la 1$ kpc) and the predicted 
luminosities of the evolving galaxies are an order of magnitude lower 
luminosities than the obtained with single-zone models. Super-solar 
metallicities are predicted to occur within $\tau_{evol} \simeq 0.5 - 0.8$ Gyrs
for these central regions (Fria\c{c}a \& Terlevich 1998; Romano et al.\,2002).

Based on the timescale for enriching the BELR gas of the 
high-redshift quasars in this study, the beginning of the first star formation
epoch can be estimated. A redshift $z \ga 4$ corresponds to an age of the 
universe of less than $\sim 1.3$\,Gyrs ($H_o = 65$\,km\,s$^{-1}$\,Mpc$^{-1}$, 
$\Omega _M = 0.3$, $\Omega _\Lambda = 0.7$). 
Assuming $\tau _{evol} \simeq 0.5 - 0.8$\,Gyrs for the star 
formation needed to enrich the gas, the first major star formation epoch 
started at $z_f \simeq 6 - 8$ for observed quasars at $z\simeq 4.5$.
It is interesting to note that this is comparable to the epoch of 
re-ionization of the universe (Haiman \& Loeb 1998; Becker et al.\,2001; 
Fan et al.\,2002).
These results are also consistent with cosmic structure formation models
(Gnedin \& Ostriker 1997) which are suggesting that the rapid assembly of 
(at least some) massive spheroidal systems, accompanied by intense star 
formation, started at $z_f \simeq 6 \,{\rm to}\,8$. Assuming that at the end
of the major star formation episode in the early 
phase of galaxy evolution, quasar activity starts in a chemical highly 
enriched environment (e.g., Granato et al.\,2001; Romano et al.\,2002) the
formation of spheroidal systems might continue until $z\simeq 2$ (Madau et 
al.\,1996; Fria\c{c}a \& Terlevich 1998; Steidel 1999).

The metallicities presented here and by Dietrich et al.\,(2002a) based on
the emission lines in quasars at $z\ga 4$ are consistent with previous 
emission line studies of $2 \la z \la 4$ quasar samples. In particular, there
is no evidence for a decline in the metallicity from $z\simeq 2$ to $z>4$
(see also Dietrich \& Hamann 2003).

In the context of galaxy evolution models, higher metallicities and shorter
evolution timescales are expected for more massive stellar systems
(Gnedin \& Ostriker 1997; Cen \& Ostriker 1999; 
Kauffmann \& Haehnelt 2000; Nolan et al.\,2001).
The close connection of quasars and the formation of massive galaxies is 
supported by the relation of the black hole mass and the mass of the 
spheroidal galaxy component (e.g., Gebhardt et al.\,2000; 
Ferrarese \& Merritt 2001).  
For elliptical galaxies, a mass -- metallicity relation has been well known 
for several decades (Sandage 1972; Faber 1973; Bica et al.\,1988).
Hence, a similar mass -- metallicity relation can be expected for quasars.
Indeed, there is some evidence for a correlation of metallicity and 
luminosity, i.e., with the black hole mass of quasars, based on broad emission
line studies similar to the analysis we present in this paper 
(Hamann \& Ferland 1993; Shemmer \& Netzer 2002; Warner et al.\,2003).

\section{Conclusion}

We investigated rest-frame ultraviolet spectra with moderate spectral 
resolution of a sample of $70$ high redshift quasars with $z \geq 3.5$.
We used emission-line flux ratios involving carbon, nitrogen, oxygen, and 
helium to estimate the metallicity of the line-emitting gas. 
To transform the observed line ratios into metallicities, we used the
results of detailed photoionization calculations as described by Hamann et 
al.\,(2002).

A comparison of the gas chemical composition derived from emission line ratios 
involving \niii ] and \nv\ indicates reasonable consistent estimates of the 
gas metallicity. The estimates of the chemical abundances based on 
\niii ]/\oiii ] 
and \nv /\ovi , \nv /(\ovi$+$\civ ), and \nv /\civ\ differ by $\sim 50$\,\%.
Based on eight individual emission line ratios we estimated an average overall
metallicity for the 70 high redshift quasars of roughly 
$Z/Z_\odot \simeq 4 \,{\rm to}\,5$.
Assuming an upper limit contribution of scattered Ly$\alpha$ emission of 
$\la 30$\,\%\ to the observed \nv $\lambda 1240$ emission 
(Hamann \& Korista 1996), the average metallicity of the BELR gas 
at high redshifts is still super-solar, within $\sim 20$\,\%\ of the
estimate above. Compared to previous studies, we find no evidence for an 
evolutionary trend in quasar metallicities from $z\simeq 2$ to $z\ga 4$.

We analyze the derived elemental abundances within the context of models 
presented by Hamann \& Ferland (1993) and Fria\c{c}a \& Terlevich (1998). With
an evolution time scale of approximately 
$\tau _{evol} \sim 0.5 ~{\rm to}~ 0.8$\,Gyrs, the epoch of the first intense
star formation is estimated to begin as early as at a redshift of 
$z_f \simeq 6 ~{\rm to}~ 8$, i.e., less than 1 Gyr of the age of the universe 
(H$_o \simeq 65$ km\,s$^{-1}$\,Mpc$^{-1}$, $\Omega _M = 0.3$, 
$\Omega _\Lambda = 0.7$).

We find a weak trend of a $L$ -- $Z/Z_\odot$ relation for the high redshift 
quasars. Due to the scatter in metallicity and the small range of covered 
luminosity the probability for a correlation by chance is $\sim 6$\,\% .
However, the trend is in good agreement with results obtained for quasars at 
lower redshift (Hamann \& Ferland 1999) and for composite spectra based on a 
large quasar sample, which we are currently investigating (Dietrich et al.\,
2002b; Warner et al.\,2003).

\begin{acknowledgements}
      We are grateful to our colleagues J.A.\,Baldwin, G.J.\,Ferland, and 
      K.T.\,Korista for helpful discussions.
      We thank Fred Chaffee and Craig Foltz for support and help with the MMT 
      observations.
      MD thanks Katherine L. Wu for comments on the manuscript.
      MD and FH acknowledge support from NASA grant NAG 5-3234 and  NSF grant 
      AST-99-84040 (University of Florida).
      MV gratefully acknowledges financial support from the Columbus 
      Fellowship.
\end{acknowledgements}

\clearpage

\figcaption[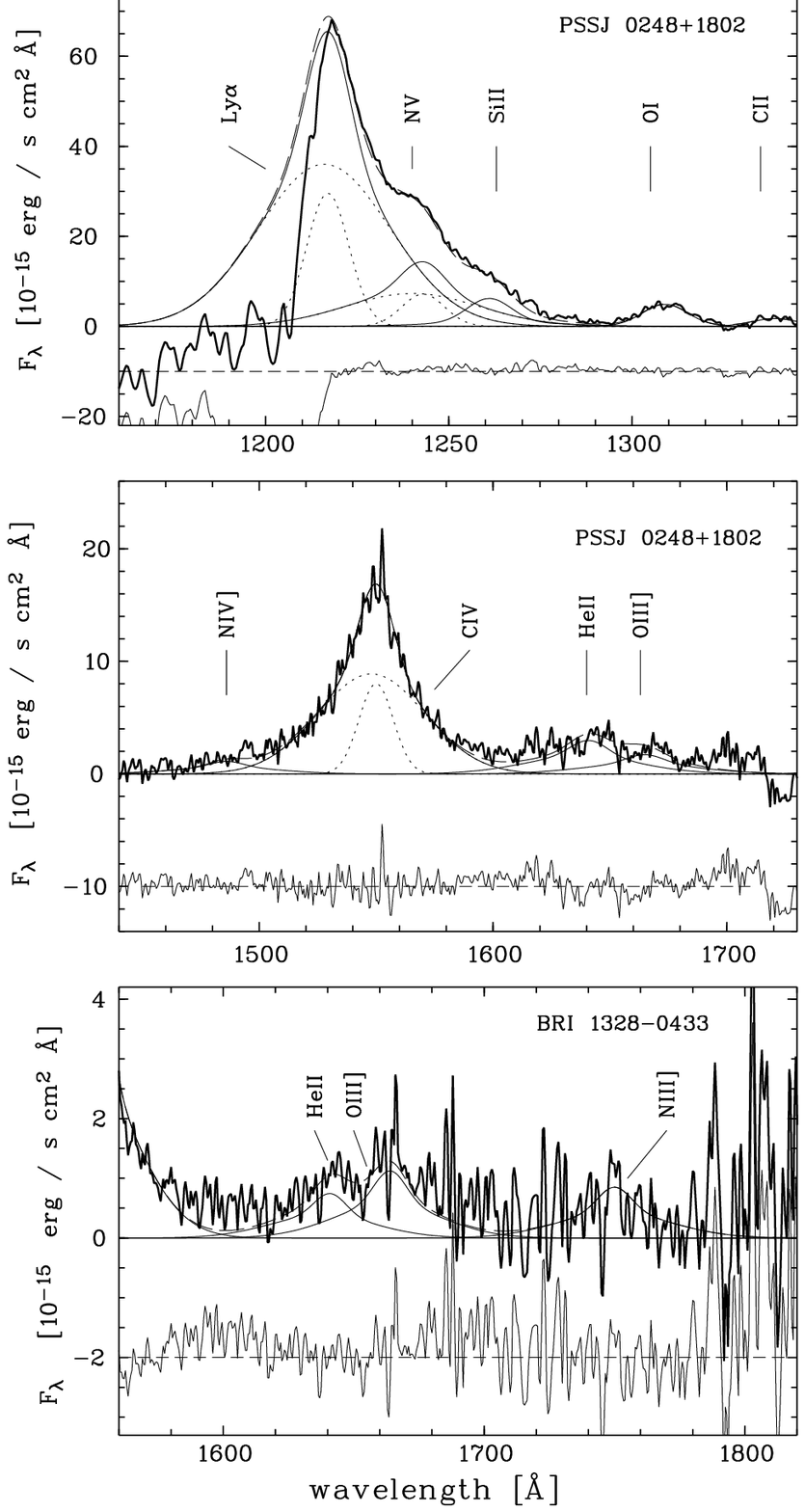]{(a) Example of the reconstruction of the Ly$\alpha $, 
            \nv $\lambda 1240$ emission line profile complex for 
            PSSJ\,0248+1802 (thick line). 
            The individual profiles are displayed as solid lines while the 
            broad and narrow components are shown as dotted lines. 
            The sum of 
            the profile fits of the individual lines is plotted as long 
            dashed line. The residuum is shown at the bottom of the figure 
            which is calculated by subtracting the individual profile fits 
            from the power-law continuum corrected quasar spectrum. The short
            dashed line indicates the zero-level for the residuum.
            (b) Example of the reconstruction of the 
            \civ $\lambda 1549$, \heii $\lambda 1640$, \oiii ]$\lambda 1663$ 
            emission line profile complex for PSSJ\,0248+1802;
            line types are the same as those shown in (a).
            (c) Example of the reconstruction of the 
            \heii $\lambda 1640$, \oiii ]$\lambda 1663$, and \niii ]$\lambda 
            1750$ emission line profile complex for BRI\,1328-0433;
            line types are the same as those shown in (a).}

\figcaption[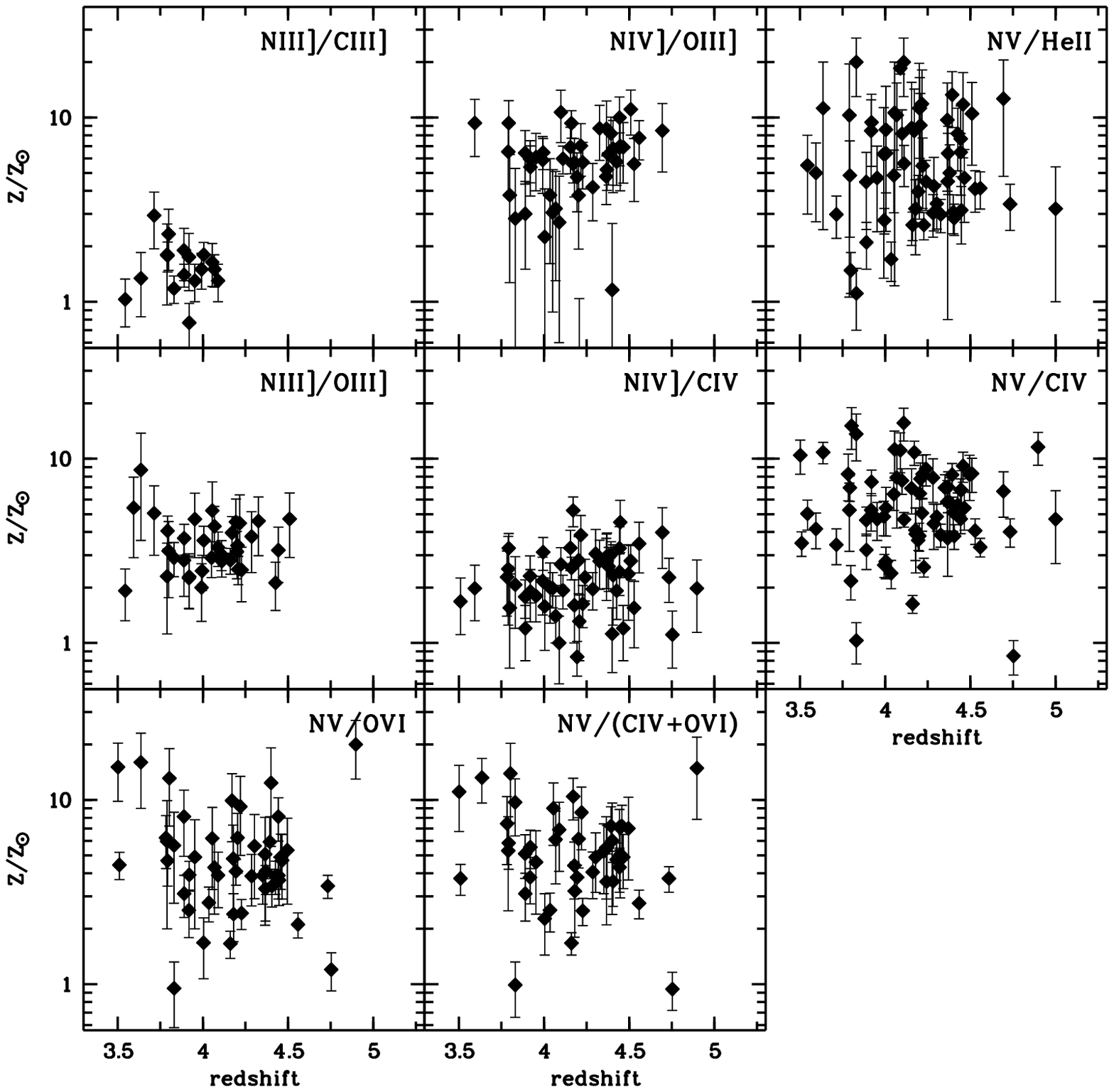]{The metallicities derived from several emission 
            line ratios are shown for the individual high redshift quasars as 
            a function of redshift.}

\figcaption[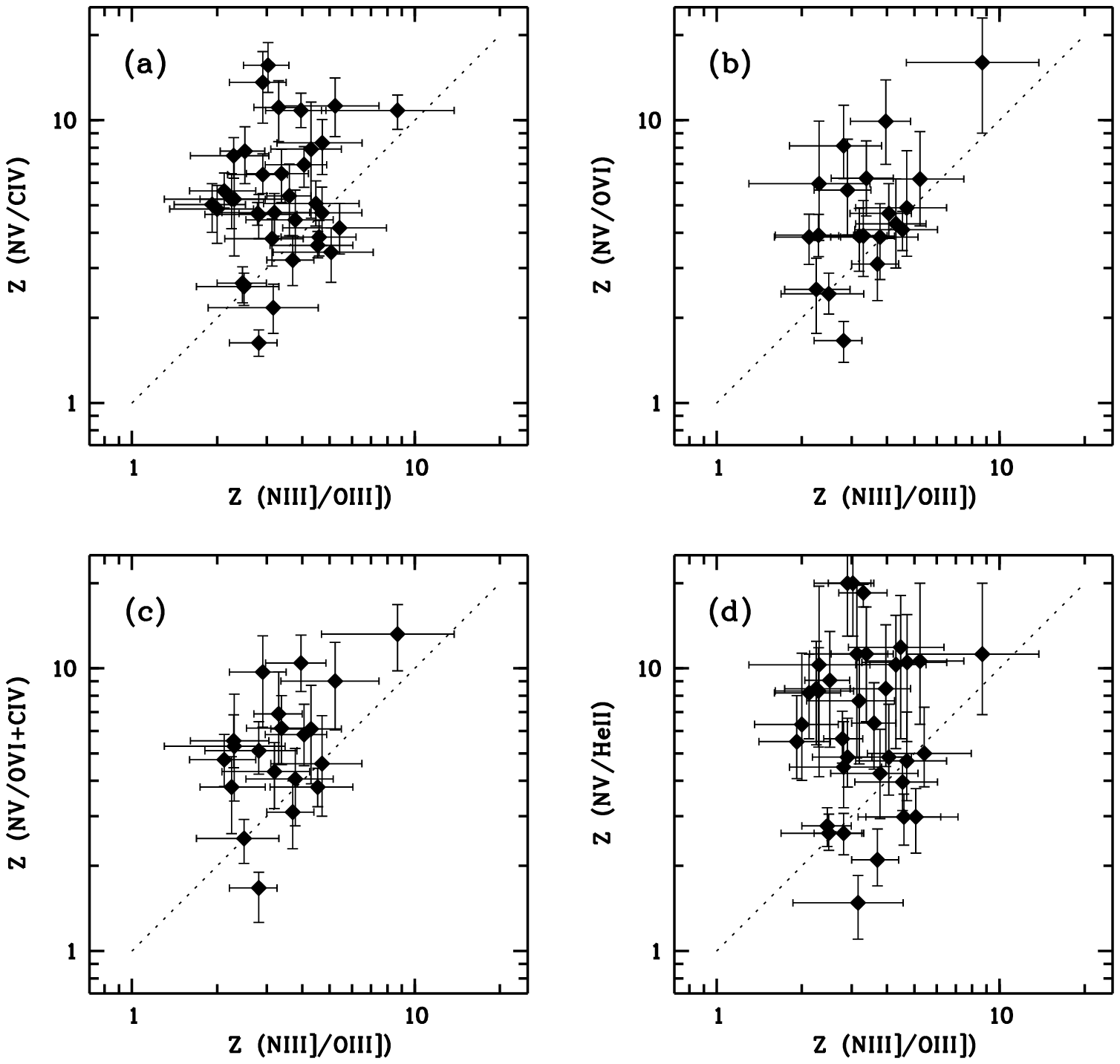]{Comparison of the metallicity estimates for each 
            quasar based on \niii ]/\oiii ] versus \nv /\civ\ (a), 
            \nv /\ovi\ (b), \nv /(\ovi $+$\civ) (c), and 
            \nv /\heii\ (d).}

\figcaption[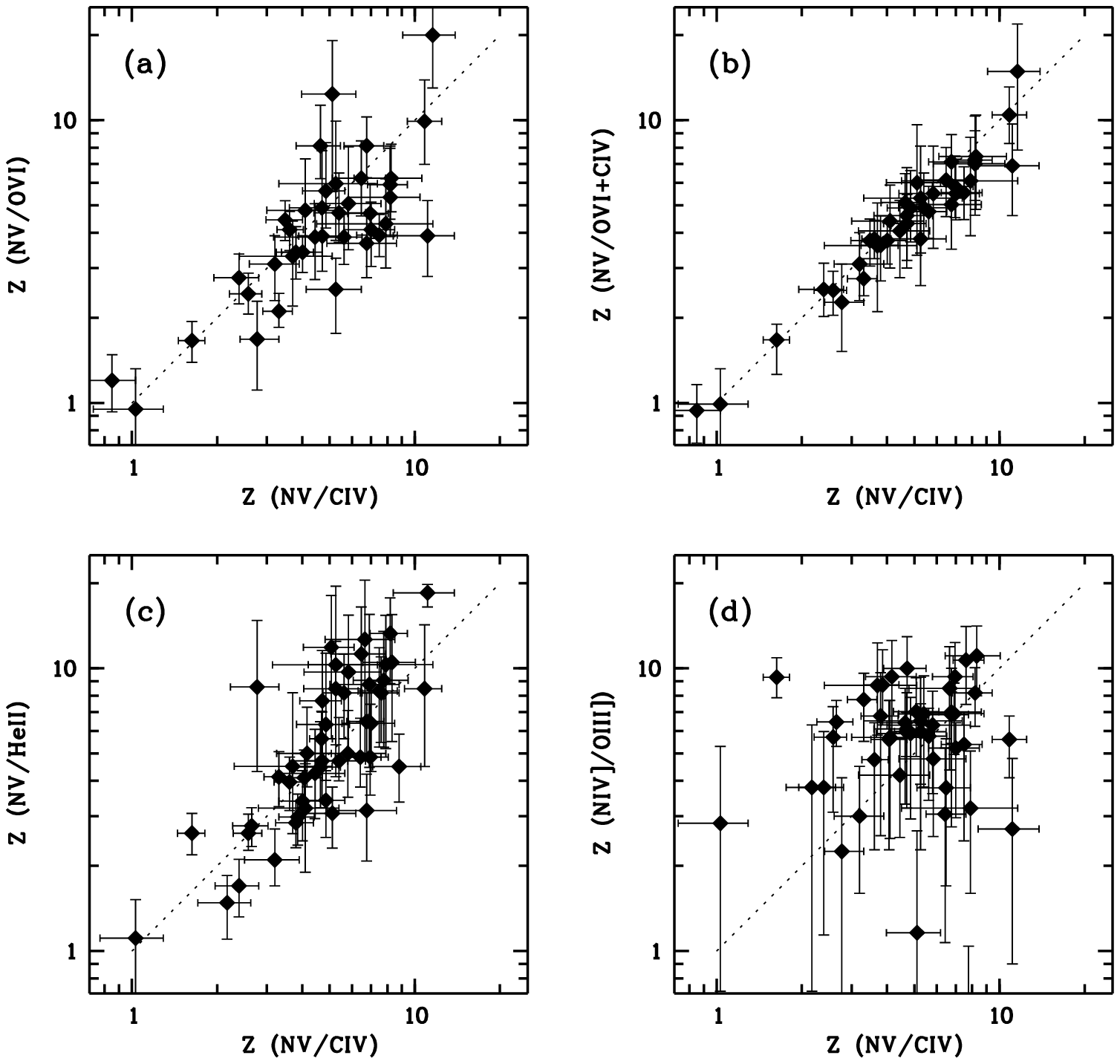]{Same as Figure 3 for
            \nv ]/\civ\ versus \nv /\ovi\ (a) , 
            \nv /(\ovi $+$\civ) (b), \nv /\heii\ (c),
            and \niv ]/\oiii ] (d).}

\figcaption[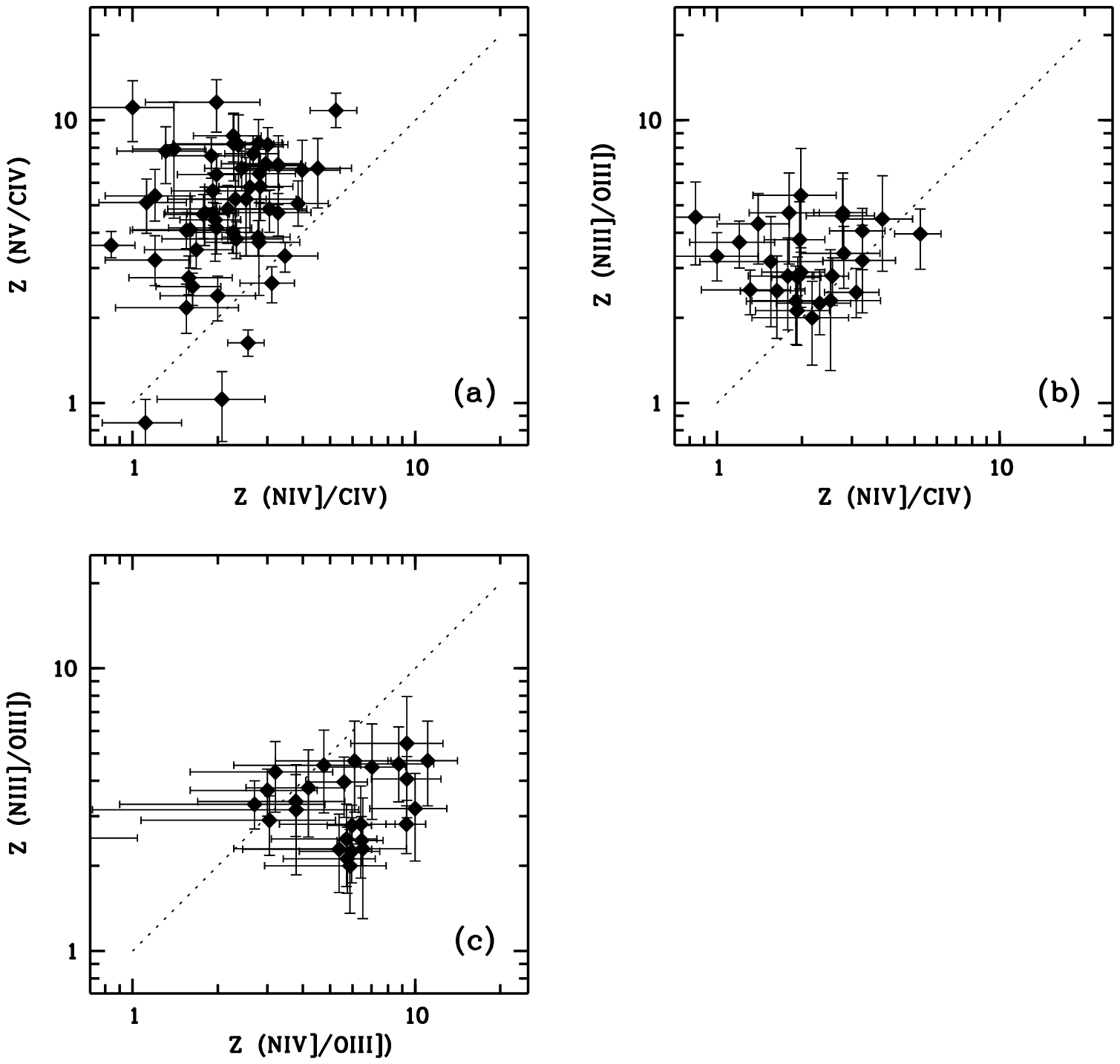]{Same as Figure 3 for
            \niv ]/\civ\ versus \nv /\civ\ (a) and 
            \niii /\oiii ] (b).
            The comparison of the metallicity given by \niv]/\oiii] versus 
            \niii ]/\oiii ] is shown in the lower left panel (c).}

\figcaption[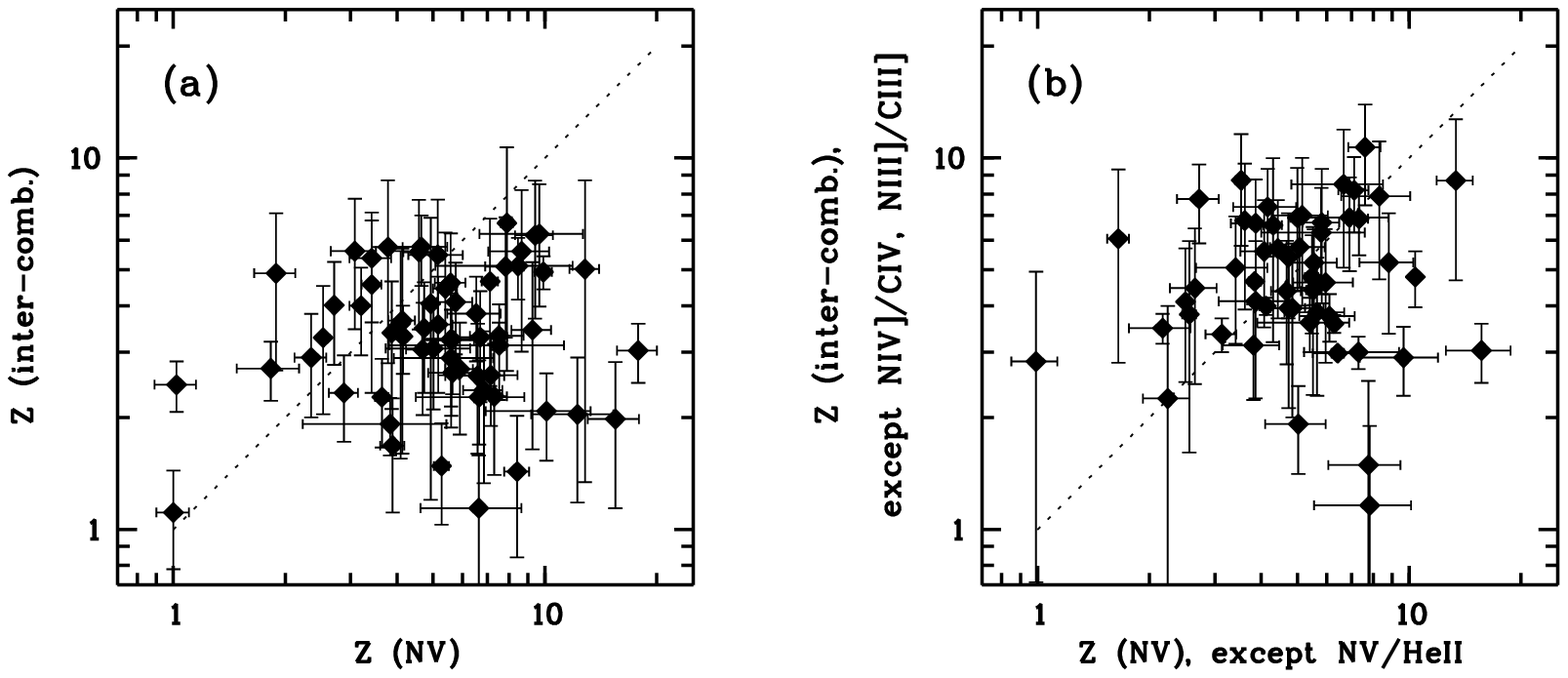]{(a) -- Comparison of the metallicities for 
            each quasar
            based on the mean metallicity obtained by the line ratios using 
            \nv\ versus the line ratios using inter-combination lines. 
            (b) -- The comparison of the mean metallicities computed 
            without using
            \nv /\heii , and \niv ]/\civ\ and \niii ]/\ciii ].}

\figcaption[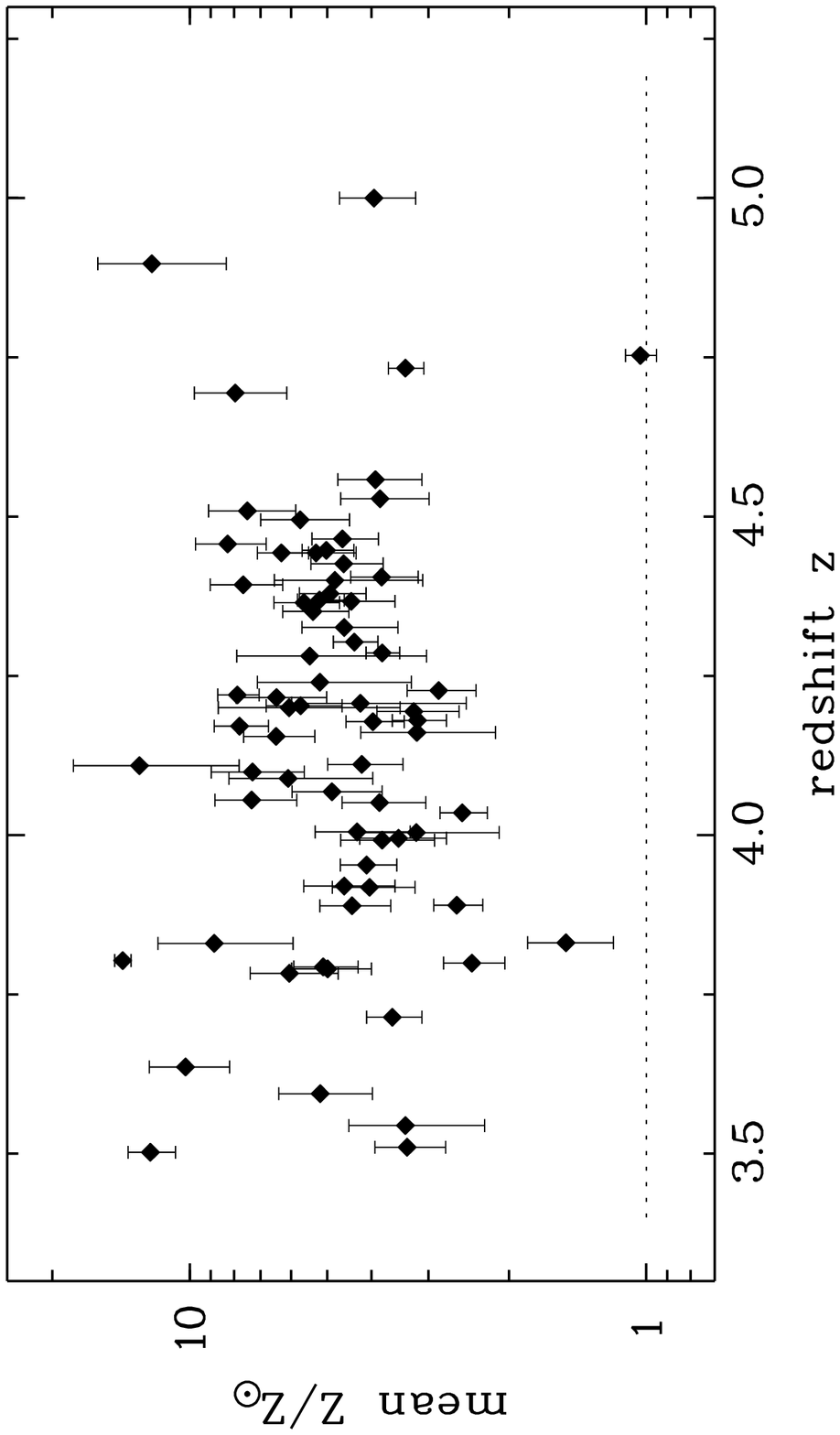]{The average metallicities of the individual high 
            redshift quasars as a function of redshift. The dotted line
            marks solar metallicity $Z_\odot$.}

\figcaption[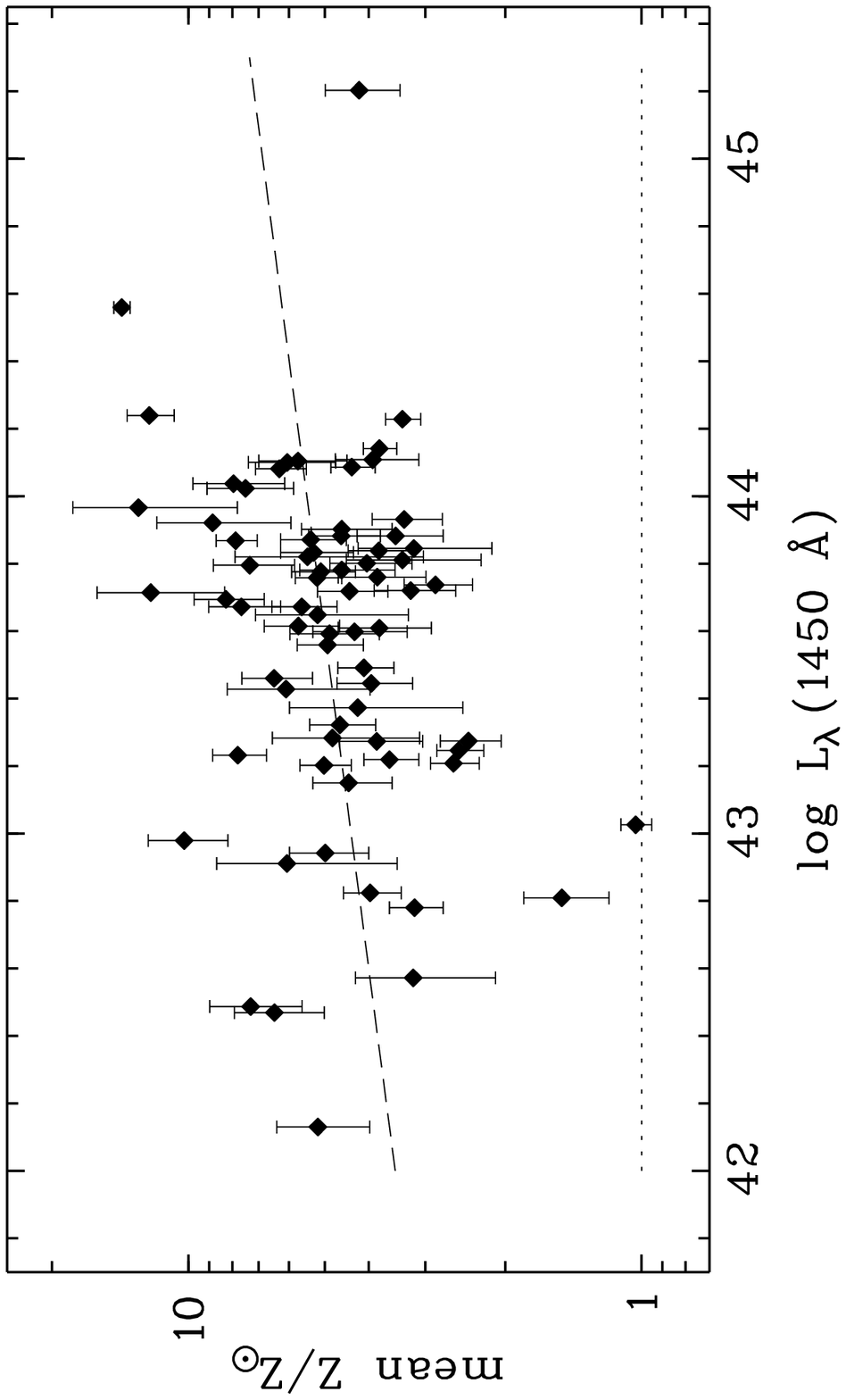]{The average metallicities of the individual high 
            redshift quasars as a function of rest-frame continuum luminosity
            $L_\lambda (1450 {\rm \AA })$. The dotted line marks solar 
            metallicity $Z_\odot$. The dashed line displays the least squares 
            fit to the measurements.}


\begin{figure}
\plotone{f1.eps}
\label{fig1a-c}
\end{figure}

\begin{figure}
\plotone{f2.eps}
\label{fig2}
\end{figure}

\begin{figure}
\plotone{f3.eps}
\label{fig3}
\end{figure}

\begin{figure}
\plotone{f4.eps}
\label{fig4}
\end{figure}

\begin{figure}
\plotone{f5.eps}
\label{fig5}
\end{figure}

\begin{figure}
\plotone{f6.eps}
\label{fig6}
\end{figure}

\begin{figure}
\plotone{f7.eps}
\label{fig7}
\end{figure}

\begin{figure}
\plotone{f8.eps}
\label{fig8}
\end{figure}

\hspace*{-15mm}
\begin{deluxetable}{lcccclcccc}
\tablewidth{0pt}
\tabletypesize{\scriptsize}
\tablecaption{The high redshift quasar sample}
\tablehead{
\colhead{quasar} &
\colhead{$\lambda \lambda $ range} &
\colhead{$z$} &
\colhead{log L$_\lambda$\tablenotemark{a}} &
\colhead{\hspace*{-3mm}ref.\tablenotemark{b}\hspace*{-3mm}} &
\colhead{quasar} &
\colhead{$\lambda \lambda $ range} &
\colhead{$z$} &
\colhead{log L$_\lambda$\tablenotemark{a}} &
\colhead{\hspace*{-3mm}ref.\tablenotemark{b}\hspace*{-3mm}}\\
\colhead{ } &
\colhead{[\AA ]} &
\colhead{ } & 
\colhead{ } &
\colhead{ } &
\colhead{ } &
\colhead{[\AA ]} &
\colhead{ } &
\colhead{ } &
\colhead{ } 
}
\startdata
\tableline
Q 0000-263    & 671 -- 1862&4.11& 45.20&\hspace*{-3mm}6  \hspace*{-3mm}&PC 1158+4635  & 788 -- 1655&4.73& 44.23&\hspace*{-3mm}6    \hspace*{-3mm}\\
PSSJ 0003+2730&1072 -- 1769&4.24& 43.65&\hspace*{-3mm}1  \hspace*{-3mm}&Q 1159+123    & 706 -- 1550&3.50& 44.24&\hspace*{-3mm}7,9  \hspace*{-3mm}\\
BR 0019-1522  & 663 -- 1773&4.53& 43.76&\hspace*{-3mm}1,4\hspace*{-3mm}&BR 1202-0725  & 631 -- 1758&4.69& 44.04&\hspace*{-3mm}1,4  \hspace*{-3mm}\\
PC 0027+0525  &1104 -- 1846&4.10& 42.49&\hspace*{-3mm}1  \hspace*{-3mm}&Q 1208+1011   & 660 -- 1600&3.80& 44.56&\hspace*{-3mm}7    \hspace*{-3mm}\\
PC 0027+0521  &1078 -- 1805&4.22& 42.47&\hspace*{-3mm}1  \hspace*{-3mm}&PC 1233+4752  & 884 -- 1744&4.45& 43.20&\hspace*{-3mm}6    \hspace*{-3mm}\\
SDSS 0032+0040&1003 -- 1663&4.75& 43.03&\hspace*{-3mm}1  \hspace*{-3mm}&PC 1247+3406  & 764 -- 1609&4.90& 43.71&\hspace*{-3mm}6    \hspace*{-3mm}\\
Q 0046-293    & 844 -- 2054&4.01& 43.60&\hspace*{-3mm}3  \hspace*{-3mm}&PKS 1251-407  & 778 -- 1707&4.47& 43.32&\hspace*{-3mm}3    \hspace*{-3mm}\\
Q 0046-282    & 763 -- 1989&3.83& 43.92&\hspace*{-3mm}2  \hspace*{-3mm}&PC 1301+4747  & 887 -- 1915&4.00& 42.57&\hspace*{-3mm}1,6  \hspace*{-3mm}\\
PSSJ 0059+0003&1056 -- 1871&4.16& 43.46&\hspace*{-3mm}1  \hspace*{-3mm}&PSSJ 1317+3531&1014 -- 1800&4.37& 43.67&\hspace*{-3mm}1    \hspace*{-3mm}\\
Q 0101-304    & 831 -- 2000&4.07& 43.59&\hspace*{-3mm}3  \hspace*{-3mm}&BRI 1328-0433 & 676 -- 1858&4.20& 43.62&\hspace*{-3mm}1,4  \hspace*{-3mm}\\
BRI 0103+0032 & 653 -- 1801&4.44& 43.83&\hspace*{-3mm}1,4\hspace*{-3mm}&Q 1330+0108   & 702 -- 1549&3.51& 43.93&\hspace*{-3mm}7    \hspace*{-3mm}\\
PC 0104+0215  & 836 -- 1835&4.17& 43.23&\hspace*{-3mm}6  \hspace*{-3mm}&APM 1335-0417 & 798 -- 1722&4.38& 43.56&\hspace*{-3mm}3,4  \hspace*{-3mm}\\
PC 0131+0120  & 900 -- 1937&3.79& 43.78&\hspace*{-3mm}1,6\hspace*{-3mm}&BRI 1346-0322 & 702 -- 1923&3.99& 43.61&\hspace*{-3mm}1,4  \hspace*{-3mm}\\
PSSJ 0152+0735&1113 -- 1863&4.05& 43.27&\hspace*{-3mm}1  \hspace*{-3mm}&Q 1422WOWccs02& 828 -- 1808&3.59& 42.13&\hspace*{-3mm}8    \hspace*{-3mm}\\
BRI 0151-0025 & 670 -- 1858&4.19& 43.72&\hspace*{-3mm}1,4\hspace*{-3mm}&PC 1450+3404  &1110 -- 1840&4.20& 42.91&\hspace*{-3mm}1    \hspace*{-3mm}\\
BRI 0241-0146 & 721 -- 1950&4.06& 43.80&\hspace*{-3mm}1,4\hspace*{-3mm}&BRI 1500+0824 & 879 -- 1942&3.95& 43.49&\hspace*{-3mm}1,3,4\hspace*{-3mm}\\
BR 0245-0608  & 711 -- 1661&4.22& 43.87&\hspace*{-3mm}4  \hspace*{-3mm}&GB 1508+5714  & 661 -- 1654&4.30& 44.09&\hspace*{-3mm}4    \hspace*{-3mm}\\
PSSJ 0248+1802&1003 -- 1778&4.44& 44.08&\hspace*{-3mm}1  \hspace*{-3mm}&PC 1548+4637  & 950 -- 2090&3.54& 43.81&\hspace*{-3mm}6    \hspace*{-3mm}\\
PC 0307+0222  & 811 -- 1747&4.39& 43.67&\hspace*{-3mm}1,6\hspace*{-3mm}&BRI 1557+0313 & 747 -- 2019&3.89& 43.21&\hspace*{-3mm}3,4  \hspace*{-3mm}\\
SDSS 0338-0021&1056 -- 1718&5.00& 43.45&\hspace*{-3mm}2  \hspace*{-3mm}&PSSJ 1618+4125&1108 -- 1836&4.21& 43.37&\hspace*{-3mm}1    \hspace*{-3mm}\\
PC 0345+0130  & 933 -- 2046&3.64& 42.98&\hspace*{-3mm}6  \hspace*{-3mm}&PC 1640+4628  & 919 -- 2016&3.71& 43.22&\hspace*{-3mm}6    \hspace*{-3mm}\\
BR 0351-1034  & 679 -- 1616&4.35& 43.87&\hspace*{-3mm}4  \hspace*{-3mm}&PC 1643+465A  & 902 -- 1950&3.79& 42.94&\hspace*{-3mm}6    \hspace*{-3mm}\\
BR 0401-1711  & 673 -- 1876&4.23& 43.74&\hspace*{-3mm}1,4\hspace*{-3mm}&PC 1640+465B  & 892 -- 1963&3.83& 42.81&\hspace*{-3mm}6    \hspace*{-3mm}\\
PC 0751+5623  & 869 -- 1794&4.28& 43.82&\hspace*{-3mm}6  \hspace*{-3mm}&GB 1745+6227  & 694 -- 1973&3.89& 43.72&\hspace*{-3mm}1,4  \hspace*{-3mm}\\
PC 0910+5625  & 876 -- 1801&4.04& 43.25&\hspace*{-3mm}6  \hspace*{-3mm}&RX 1759.4+6638&1083 -- 1756&4.33& 43.78&\hspace*{-3mm}1    \hspace*{-3mm}\\
BR 0951-0450  & 661 -- 1833&4.37& 43.76&\hspace*{-3mm}1,4\hspace*{-3mm}&Q 2000-330    & 703 -- 1574&3.78& 44.10&\hspace*{-3mm}5    \hspace*{-3mm}\\
BRI 0952-0115 & 681 -- 1776&4.43& 43.88&\hspace*{-3mm}1,4\hspace*{-3mm}&PC 2047+0123  & 902 -- 1978&3.80& 43.27&\hspace*{-3mm}6    \hspace*{-3mm}\\
PC 0953+4749  & 866 -- 1743&4.46& 43.69&\hspace*{-3mm}6  \hspace*{-3mm}&Q 2133-4311   & 833 -- 1798&4.18& 42.82&\hspace*{-3mm}3    \hspace*{-3mm}\\
BRI 1013+0035 & 660 -- 1785&4.41& 43.84&\hspace*{-3mm}1,4\hspace*{-3mm}&Q 2133-4625   & 829 -- 1974&4.18& 42.78&\hspace*{-3mm}3    \hspace*{-3mm}\\
BR 1033-0327  & 636 -- 1782&4.51& 44.02&\hspace*{-3mm}1,4\hspace*{-3mm}&Q 2134-4521   & 805 -- 1732&4.37& 43.15&\hspace*{-3mm}3    \hspace*{-3mm}\\
BRI 1050-0000 & 666 -- 1784&4.29& 44.14&\hspace*{-3mm}1,4\hspace*{-3mm}&Q 2203+292    & 865 -- 1770&4.40& 43.28&\hspace*{-3mm}1,6  \hspace*{-3mm}\\
PSSJ 1057+4555&1039 -- 1865&4.11& 43.97&\hspace*{-3mm}1  \hspace*{-3mm}&BR 2212-1626  & 706 -- 1932&4.00& 43.88&\hspace*{-3mm}1,4  \hspace*{-3mm}\\
BRI 1108-0747 & 734 -- 1938&3.92& 43.90&\hspace*{-3mm}1,4\hspace*{-3mm}&BR 2237-0607  & 682 -- 1758&4.56& 44.18&\hspace*{-3mm}1,3,4\hspace*{-3mm}\\
BRI 1110+0106 & 714 -- 1957&3.92& 43.80&\hspace*{-3mm}1,4\hspace*{-3mm}&BR 2248-1242  & 706 -- 1899&4.16& 43.85&\hspace*{-3mm}1,4  \hspace*{-3mm}\\
BRI 1114-0822 & 654 -- 1598&4.50& 44.10&\hspace*{-3mm}4  \hspace*{-3mm}&PC 2331+0216  & 855 -- 1944&4.09& 43.43&\hspace*{-3mm}1,3,6\hspace*{-3mm}\\
\enddata
\tablenotetext{a}{log\,$L_\lambda$ refers to the continuum luminosity 
                  $L_\lambda (1450 {\rm \AA })$ in erg\,s$^{-1}$\,\AA $^{-1}$}
\tablenotetext{b}{1 - Constantin et al.\,2002;
                 2 - Dietrich et al.\,1999;
                 3 - Dietrich et al.\,2002a;
                 4 - Storrie-Lombardi et al.\,1996;
                 5 - Sargent et al.\,1989;
                 6 - Schneider et al.\,1991a,b;
                 7 - Steidel \& Sargent, priv.comm.;
                 8 - Steidel, priv.comm.; and
                 9 - Sargent et al.\,1988}
\end{deluxetable}

\clearpage
\begin{deluxetable}{lcccc}
\tablewidth{0pt}
\tablecaption{Mean metallicities based on the individual emission line ratios.
              The number of quasars for which the individual line ratios are 
              available is given in column (2). The mean metallicities we 
              derived for the scattered light corrected \nv $\lambda 1240$ and
              \civ $\lambda 1549$ emission are given in column (5).}
\tablehead{
\colhead{ratio} &
\colhead{n}     &
\colhead{mean}  &
\colhead{median}&
\colhead{mean$_{corr}$} \\
\colhead{(1)}   &
\colhead{(2)}   &
\colhead{(3)}   &
\colhead{(4)}   & 
\colhead{(5)}   
}
\startdata
\niii ]$\lambda 1750$/\ciii ]$\lambda 1909$&17&$1.6\pm0.2$&1.5&\nodata \\
\niii ]$\lambda 1750$/\oiii ]$\lambda 1663$&35&$3.6\pm0.3$&3.2&\nodata \\
\niv ]$\lambda 1486$/\civ $\lambda 1549$   &54&$2.3\pm0.2$&2.2&$2.6\pm0.2$\\
\niv ]$\lambda 1486$/\oiii ]$\lambda 1663$ &46&$6.0\pm0.5$&6.0&\nodata \\
\nv $\lambda 1240$/\heii $\lambda 1640$    &61&$6.8\pm0.8$&5.5&$4.0\pm0.6$\\
\nv $\lambda 1240$/\civ $\lambda 1549$     &70&$6.1\pm0.4$&5.3&$4.6\pm0.4$\\
\nv $\lambda 1240$/\ovi $\lambda 1034$     &47&$5.6\pm0.7$&4.4&$3.5\pm0.5$\\
\nv $\lambda 1240$/(\ovi$+$\civ )          &47&$5.7\pm0.6$&5.1&$4.0\pm0.5$\\
\tableline				       
all ratios                                 &70&$5.3\pm0.3$&4.7&\nodata \\
inter-combination lines                    &62&$3.6\pm0.3$&3.3&\nodata \\
\nv -ratios                                &70&$6.2\pm0.4$&5.6&\nodata \\
\enddata
\end{deluxetable}

\end{document}